\documentstyle[aps,prb,eqsecnum,epsf,preprint,tighten]{revtex}
\begin{document}
\title{Boundary Critical Phenomena in SU(3) ``Spin'' Chains}
\author{Ian Affleck$^{1}$, Masaki Oshikawa$^2$ and Hubert
Saleur$^3$}
\address{$^1$Canadian Institute for Advanced
Research  and Department of Physics and Astronomy, University of
British
Columbia,
Vancouver, B.C., Canada, V6T 1Z1 \\
$^2$Department of Physics, Tokyo Institute of Technology, \\ Oh-
okayama,
Meguro-ku,
 Tokyo 152-8551, Japan\\
$^3$ Department of Physics, University of Southern California, Los
Angeles,
CA90089-
0484, USA}
\maketitle
\begin{abstract}
 SU(3)-invariant ``spin'' chains with a single impurity, such as
a  modified exchange coupling on one link, are analyzed using
boundary conformal field theory techniques.  These chains are
equivalent to a special case of the ``$tJV$'' model, i.e. the $tJ$
model with a nearest neighbour repulsion added.  In the continuum
limit they are equivalent to two free bosons at a special value of
the compactification radii.  The SU(3) symmetry, which is made
explicit in this formulation, provides insight into the exact
solution of a non-trivial boundary critical point found earlier in
another formulation of this model as a theory of quantum Brownian motion. 
\end{abstract}
\section{Introduction}
Recently, there has been considerable interest in conformal field theory (CFT)
with boundaries in the context of open string theory, classical
statistical mechanics and quantum impurity problems in condensed matter
physics.  Of particular interest are certain non-trivial boundary 
critical points
first discovered\cite{Furusaki,Kane}  in the context of a constriction
in a quantum wire.  While the continuum limit of these models is simply
two free bosons (one for charge and one for spin), the non-trivial
boundary critical points {\it do not} correspond to any variant of simple
Dirichlet (D) and Neumann (N) boundary conditions.  They
correspond to boundary critical points with intermediate (neither 0 nor 1)
transmission amplitude through the constriction.  The original approach
of Kane and Fisher\cite{Kane}  was only able to study
them using a type of $\epsilon$-expansion around limiting values
of the  compactification radii or bulk interaction parameters where they
became trivial.  Later Yi and Kane\cite{Yi} reinvestigated these critical
points in the context of a model of quantum Brownian motion on
a triangular lattice finding a special value of compactification radii
where the non-trivial critical point could be solved exactly.  This
special point was very recently investigated by the present
authors\cite{Affleck1}
using the CFT techniques of conformal embedding and fusion, relating
it to the 3-state Potts model with a boundary.\cite{Affleck2}  However, a
general solution for these non-trivial boundary critical points, for all
values of the compactification radii where they occur is still lacking.
More generally, a framework seems to be lacking for understanding
non-trivial boundary critical points in multi-component free boson theories.

 The purpose of the present work is to provide yet another view of
the special soluble point.   In this case the microscopic formulation
is an SU(3) ``spin'' chain with the objects transforming under
the fundamental representation of SU(3) at each lattice site and
a permutation Hamiltonian.  We will show that the $SU(3)$ symmetry
is sufficient to uniquely pick out the solvable non-trivial critical
point without any fine-tuning.
 It is convenient to introduce 3 fermion
annihilation operators, $\eta_{j\alpha}$ on each site, $j$ with
$\alpha = 0,1,2$ and a single occupancy constraint:
\begin{equation}
\eta^{\dagger \alpha}_j\eta_{j\alpha}=1.\label{constraint}\end{equation}
We use a superscript for annihilation operators and repeated
indices (one upper and one lower) are always summed.  The permutation
Hamiltonian can then be written:
\begin{equation}
H=(1/2)\sum_jJ_j\eta^{\dagger \alpha}_j\eta_{\beta j}\eta^{\dagger \beta}_{j+1}
\eta_{j+1\alpha}.\label{Ham1}\end{equation}
This is the 3-component Lai-Sutherland model\cite{Lai} in the case where all
three objects obey fermionic statistics.  (The same model is obtained
if all three objects obey bosonic statistics.)  This model is equivalent
to a special case of the ``$tJV$'' model, i.e. the $tJ$ model with an
additional nearest neighbour repulsion, as we review in Sec. 4.  This
model is Bethe ansatz integrable and has a gapless excitation spectrum.
Its continuum limit\cite{Affleck4} is the $SU(3)_1$ Wess-Zumino-Witten (WZW)
non-linear $\sigma$-model.  We will study this model in the case
where one or more links have a modified exchange coupling $J_j$. All
other links have a fixed antiferromagnetic exchange coupling $J>0$.  The
behaviour is quite different than that of the corresponding $SU(2)$
spin chains (with S=1/2). In the $SU(2)$ case modifying one link
 produces a renormalization group (RG) flow to an open chain fixed
point, corresponding to the exchange coupling on the modified link
renormalizing to 0 (or $\infty$).  However, in the $SU(3)$ case
a flow instead occurs to a non-trivial fixed point which does
not correspond to an open chain nor to a uniform chain.  This
corresponds to the intermediate transmission coefficient fixed
point\cite{Furusaki,Kane} in the $tJV$ formulation.    The $SU(3)$
symmetry of the model makes it possible to study the fixed point
using fusion.  Our approach is to first regard the right-movers
as a second branch of left-movers, reflecting them at the impurity
location.  The two copies of left-moving $SU(3)_1$ WZW excitations
can then be represented by the conformal embedding:
\begin{equation}
SU(3)_1\times SU(3)_1 \equiv SU(3)_2\times \hbox{Potts}.
\end{equation}
This corresponds to the sum of central charges:
\begin{equation}
2+2=16/5+4/5.\label{2+2}\end{equation} The non-trivial critical
point can be reached by fusion either in the $SU(3)_2$ or Potts
sector. We note that the original solution of this model by Yi and
Kane\cite{Yi} mapped it onto the
 3-channel $SU(2)$ Kondo problem, corresponding to the
$SU(2)_3$ WZW model.  This model is related by a duality
transformation to $SU(3)_2$ WZW model.  We show that the
spinful Luttinger liquid model at the
value of the bulk interaction parameters, $g_{\sigma}=2$,
$g_{\rho}=2/3$, where the non-trivial critical point can be
studied exactly, has an $SU(3)$ symmetry.  This provides
the most natural understanding of what is special about
this point in parameter space.

In the next section we review the analogous boundary critical
phenomena in the ordinary (S=1/2) SU(2) chain.  Section III
contains our new results on the SU(3) chain.  Sec. IV discusses
the connection with the $tJV$ model.  Sec. V discusses more general 
models where the $SU(3)$ symmetry is broken down to $SU(2)\times U(1)$, 
either at the boundary only, or also in the bulk.  In Sec. VI we discuss 
the connection of the $SU(3)$ spin chain boundary critical 
behaviour with that which occurs in the 2-channel $SU(3)$ Kondo 
model and in the triangular lattice quantum Brownian motion model.  We 
also comment on the extension of this work to the general $SU(N)$ case. 

\section{SU(2) Case}
Here we consider an ordinary SU(2) S=1/2 chain with a single impurity.
The Hamiltonian is written:
\begin{equation}
H=\sum_jJ_j\vec S_j\cdot \vec S_{j+1}.\end{equation}
The exchange couplings, $J_j=J$ on all links except for one where
$J_0=J'$ or two neighbouring links where $J_0=J_1=J'$.  It was
argued in Ref. (\onlinecite{Eggert}) that the only fixed points that occur in
this problem correspond to the uniform and open chain.  Modifying
one link leads to a flow to the open chain fixed point.  If $J'<J$,
we may think of $J'$ as renormalizing to $0$, corresponding to an
open chain.  If $J'>J$, we may think of $J'$ as renormalizing
to $\infty$.  In this limit the two spins at sites $0$ and $1$ form
a singlet and decouple from the rest of the spins which therefore
correspond again to an open chain.  Thus $J'=J$ represents
an unstable fixed point whereas $J'=0$ or $\infty$ are stable
fixed points.  This conjecture was based on an analysis of
the operator content at the uniform and open fixed points.
This can be conveniently performed using non-abelian
bosonization.\cite{Affleck5}  The spin operators, in the continuum limit
are represented in terms of the fundamental field $g^\alpha_\beta$,
and currents, $\vec J_{L,R}$
of the $SU(2)_1$ WZW model as:
\begin{equation}
\vec S_j\approx (\vec J_L+\vec J_R)+\hbox{constant}(-1)^jtr (g\vec \sigma ).
\end{equation}
By using the operator product expansion (OPE) one can show that
\begin{equation}
\vec S_j\cdot \vec S_{j+1} \approx \hbox{constant}(-1)^jtr g
+\hbox{constant}\vec J_L\cdot \vec J_R.\label{SSbos}
\end{equation}
Thus the modified link corresponds to a local interaction at the
origin, in the low energy effective Hamiltonian of the form:
\begin{equation}
\delta H \propto (J'-J)tr g (0).\end{equation} Since $g$ has
scaling dimension $1/2$, this is relevant.  (Recall that
interactions occurring at only one point are relevant if they have
dimension $<1$.)  To check the stability of the open fixed point
we must consider its boundary operator content.  Boundary
operators are contained in the {\it chiral} part of the  $SU(2)$
WZW theory and, in this case just correspond to the identity
conformal tower.  Thus the lowest dimension operator corresponding
to the spin at the end of an open chain is the current, of
dimension $1$.  Coupling the two boundary spins together across
the open link gives an operator of dimension $2$ which is
irrelevant. This conjecture is thus shown to be consistent with
the stability of the open and uniform chain fixed points.  The
conjecture was further tested by numerical work.\cite{Eggert}  The
situation is quite different for two neighbouring modified links.
In that case the uniform chain fixed point is stable and the open
chain fixed point is unstable.  The crucial difference at the
uniform chain fixed point is that the relevant operator $tr g$
cancels due to the $(-1)^j$ factor in Eq. (\ref{SSbos}) leaving
only the irrelevant operator $d tr g/dx$ of dimension $3/2$.  On
the other hand, if we consider the limit $J'\to 0$ on two links,
we obtain two open chains and one decoupled spin.  The RG
equations in this case are the same as in the 2-channel S=1/2
Kondo problem (the two channels corresponding to the left and
right side of the impurity spin).  An infinitesimal positive $J'$
is marginally relevant.  In this case we think of the perturbed
chain as ``healing''.  The effects of the local perturbation
disappear in the low energy effective Hamiltonian.

These two fixed points and the various RG flows between them
can also be studied by the fusion technique,\cite{Affleck3} which played an
essential role in the CFT study of the multi-channel Kondo problem.
This gives a way of determining new boundary critical points
from a starting reference critical point.  In some cases this
leads to the discovery of new critical points or a possible proof
of the absence of additional critical points given certain
completeness assumptions.

A starting point for fusion is to regard the right-movers as a
second branch of left movers.  This is possible because left and
right movers are, in a sense, decoupled in the conformal field
theory; it is only the impurity interactions which couple them
together.  (We note that  true boundary models, such as a spin
chain on a semi infinite line with interactions near the origin,
can also be formulated entirely in terms of left movers on the
infinite line.  However, in this case, no doubling of the number
of degrees of freedom occurs.)
  In the spin chain problem
we thus obtain a model with two flavors of left-moving WZW excitations,
 $SU(2)_1\times SU(2)_1$,
defined on an infinite line with the impurity interactions at the
origin. It turns out\cite{Affleck3} that to study the fixed points
using fusion it appears necessary to then use a conformal
embedding:
\begin{equation}
SU(2)_1\times SU(2)_1 = SU(2)_2\times \hbox{Ising}.\end{equation}
The $SU(2)_2$ excitations carry the diagonal $SU(2)$ quantum numbers and
the Ising $Z_2$ symmetry corresponds to switching the two $SU(2)_1$ groups,
or equivalently a parity transformation in the original formulation.  The
central charge adds up correctly, recalling that $c=1$, $3/2$ and $1/2$ for
$SU(2)_1$, $SU(2)_2$ and Ising respectively.  The original non-chiral
WZW fields at $x=0$ can now be represented as:
\begin{eqnarray}
\vec J_L+\vec J_R &=& \vec J\nonumber \\
tr g\vec \sigma &\propto& \vec \phi \nonumber \\
tr g &\propto & \epsilon .\label{SU2bos}\end{eqnarray}
Here $\vec J$ is the (chiral) current operator in $SU(2)_2$,
$\vec \phi$ is the spin-1 primary field of dimension $1/2$ and
$\epsilon$ is the energy operator of the Ising model, also
of dimension $1/2$. Note that these are all chiral operators and
the dimensions of $\epsilon$ and $g$  are
 1/2 of the dimensions of the corresponding
scalar operators in the bulk Ising or WZW models (corresponding
to the ``left-moving parts'').  In particular, in
the case of the Ising model, we could think of $\epsilon$ as
corresponding to the chiral Majorana fermion field.
 The next step is to represent the
various partition functions that occur with either open or uniform
b.c.'s at both ends, for a finite system of length $l$, in
terms of this conformal embedding.  We think of the system as
consisting of two sections of chain, both of length $l$, which
may either be joined together, or separated at their two ends,
at $x=0$ and $l$.  Thus the uniform-uniform system is a periodic
chain of length $2l$, the uniform-open system is a single open
chain of length $2l$ and the open-open system is two open chains,
both of length $l$.  We must also keep track of whether the
number of microscopic $S=1/2$ operators  is even or odd, giving
a total of 7 different partition functions.  We express these
partition functions, at temperature $\beta^{-1}$, in terms
of the modular parameter:
\begin{equation} q\equiv e^{-\pi \beta /l}.\end{equation}
We write these partition functions in terms of the (chiral)
characters $\chi_j^{(1)}$ of the $SU(2)_1$ model for $j=0$ or
$1/2$, $\chi_j^{(2)}$ of the $SU(2)_2$ model for $j=0$, $1/2$ or
$1$ and $\chi_j^{I}$ of the Ising model for $j=0$, $1/2$ and $1$.
In the Ising case we have labeled identity operator, order
parameter and energy operator in terms of a parameter $j=0$, $1/2$
and $1$ respectively. This is appropriate due to an isomorphism of
the fusion rule coefficients and modular S-matrix between the
$SU(2)_2$ and Ising CFT's. We also carefully take into account the
universal $1/l$ terms in the groundstate energy which are $-\pi
/12l$ for a periodic chain of length $2l$ and $-\pi /24l$ for a
(single) open chain of length $l$.    We thus obtain the following
partition functions, written first in terms of $SU(2)_1\times
SU(2)_1$ characters and then in terms of $SU(2)_2\times$Ising
characters:
\begin{eqnarray}
q^{1/12}Z^e_{UU}(q)&=&\left[\chi_0^{(1)}(q)\right]^2
+\left[ \chi_0^{(1/2)}(q)\right]^2=\left[ \chi_0^{(2)}(q)+\chi_1^{(2)}(q)\right]
\left[\chi_0^I(q)+\chi_1^I(q)\right]
\nonumber \\
q^{1/12}Z^o_{UU}(q)&=&2\chi_0^{(1)}(q)\chi_{1/2}^{(0)}(q)=2\chi_{1/2}^{(2)}(q)
\chi_{1/2}^I(q)
\nonumber \\
q^{1/12}Z^e_{UO}(q)&=&q^{1/16}\chi_0^{(1)}(\sqrt{q})=\left[ \chi_{0}^{(2)}(q)
+\chi_{1}^{(2)}(q)\right] \chi_{1/2}^I(q)
\nonumber \\
q^{1/12}Z^o_{UO}(q)&=&q^{1/16}\chi_{1/2}^{(1)}(\sqrt{q})=\chi_{1/2}^{(2)}(q)
\left[  \chi_{0}^I(q)+\chi_{1}^I(q)\right]
\nonumber \\
q^{1/12}Z^{ee}_{OO}(q)&=&\left[ \chi_{0}^{(1)}(q)\right]^2=\chi_{0}^{(2)}(q)
\chi_{0}^I(q)+\chi_{1}^{(2)}(q)
\chi_{1}^I(q)\nonumber \\
q^{1/12}Z^{eo}_{OO}(q)&=&\chi_{1/2}^{(1)}(q)\chi_{0}^{(1)}(q)=\chi_{1/2}^{(2)}(q)
\chi_{1/2}^I(q)
\nonumber \\
q^{1/12}Z^{oo}_{OO}(q)&=&\left[ \chi_{1/2}^{(1)}(q)\right]^2=\chi_{0}^{(2)}(q)
\chi_{1}^I(q)+\chi_{1}^{(2)}(q)
\chi_{0}^I(q)
\end{eqnarray}
Here the superscripts denote even or odd length chains and the lower subscripts
denote uniform or open b.c.'s.
Now we use the fusion rules, which are isomorphic for $SU(2)_2$ and Ising.  These
are:
\begin{eqnarray}
1/2 \times 1/2 &=& 0+1\nonumber \\
1 \times 1/2 &=& 1/2 \nonumber \\
1\times 1 &=& 0.\end{eqnarray} We can now check that fusion
correctly takes us between the various partition functions in a
way which corresponds to the various RG flows.  For instance,
suppose we start with $Z_{OO}^{ee}$, open-open b.c.'s with an even
number of spins in each chain.  Now consider adding one extra spin
at $x=0$ which is weakly (and symmetrically) coupled to both
chains.  As discussed above, this induces a Kondo-type RG flow to
the uniform-open fixed point, now with an odd number of sites.
Since we have induced this flow by coupling to an S=1/2 impurity
it is natural to associate this flow with fusion with S=1/2.  In
fact, this is exactly what occurs in the Kondo problem. Now
applying the fusion rules to $Z^{ee}_{OO}$, we see that both
characters $\chi_0^{(2)}$ and $\chi_1^{(2)}$ get replaced by
$\chi_{1/2}^{(2)}$ thus turning $Z^{ee}_{OO}$ into $Z^{o}_{UO}$.
Similarly fusion turns $Z^{oo}_{OO}$ into $Z^{o}_{UO}$ and
$Z^{e0}_{OO}$ into $Z^{e}_{UO}$. We now add another spin at $x=l$,
coupled to both chain ends and thus inducing a Kondo-type flow to
the UU fixed points.  Again it can be checked that fusion turns
$Z^e_{UO}$ into $Z^o_{UU}$ and $Z^o_{UO}$ into $Z^e_{UU}$.  It is
also interesting to start with the OO case and consider fusion
with the j=1 primary.  In this case we see that $Z^{ee}_{OO}$ is
interchanged with $Z^{oo}_{OO}$ while $Z^{eo}_{OO}$ goes into
itself.  The associated RG flow now corresponds to introducing an
$S=1$ impurity at $x=0$ or two $S=1/2$'s with a ferromagnetic
coupling.  The flow back to the open fixed point corresponds to
the $S=1$ impurity being screened by one $S=1/2$ spin from the end
of each chain.  Alternatively, if we have two weakly
ferromagnetically coupled $S=1/2$ impurities, we may think of one
of them attaching onto the end of each chain and asymptotically
decoupling from each other.  In either picture we end up getting
the flow obtained from fusion.  Similarly, $Z^e_{UO}$ and
$Z^o_{UO}$ go into themselves under fusion with j=1.

We may also consider fusion in the Ising sector. Note that the OO
partition functions simply map into themselves under fusion with
 $\epsilon$ but map into the OU partition functions under fusion
with $\sigma$.  The corresponding RG flows correspond to the
two-impurity Kondo problem.  We can imagine adding two S=1/2
impurities at $x=0$ which couple to each other with strength $J''$
and also couple one to each chain with strength $J'$.  The stable
fixed points for this problem are just the open chain. If $J''$ is
too large compared to $J'$ the two impurities just form a singlet
and decouple.  If $J''$ is too small then one impurity can couple
onto the end of each chain, the chains remaining open. However if
the ratio of $J''$ to $J'$ is just right then the system can heal,
flowing to the uniform fixed point.  These three cases correspond
to the inter-impurity singlet, Kondo screened and non-Fermi liquid
fixed points in the two-impurity Kondo problem respectively. The
second case (independent impurity screening by the chains)
corresponds to fusion with $\epsilon$ while the uniform chain
corresponds to fusion with $\sigma$.  Again the same type of
fusion was used in the CFT treatment of the two impurity Kondo
problem.  We also note that the same fusion process describes the
effect of adding two more impurities at $x=l$ to the UO chain.
Fusion with $\epsilon$ corresponds to attaching one spin to each
open chain but fusion with $\sigma$ corresponds to the defect
healing, producing a flow to the uniform fixed point.

The Ising symmetry may be given a physical interpretation.  The $Z_2$
symmetry corresponds to parity, reflection around the origin.
One was of seeing this is to note that the boundary operator introduced
by a single modified link, $tr g(0)$, corresponds to $\epsilon$
after our conformal embedding and this operator corresponds to
a boundary magnetic field in the Ising model.\cite{Cardy}
  A single
modified link breaks this symmetry explicitly, whereas two equally
modified links [between (-1) and 0 and between 0 and 1] do not.
The analogue of Ising order in the S=1/2 chain is a spontaneously
dimerized state.  This does not occur for the Heisenberg model,
although it does with sufficiently strong next nearest neighbour
interaction, as in the Majumdar-Ghosh model.\cite{Majumdar}  The
quasi-long range dimer-dimer correlation function indicates that
the Heisenberg model is in a critical state with respect to this
type of order.  Strengthening the coupling on a single link
locally favors one of the two dimer states.  It is like applying a
boundary magnetic field to the critical Ising model.  This is a
relevant perturbation and in the infrared limit is like applying a
spin up boundary condition.  Thus the flow from uniform to open
fixed points in the spin chain corresponds to the flow from free
to fixed boundary conditions in the Ising model.  In both cases
the model is responding to a symmetry breaking perturbation acting
only at the boundary.

The fact that no new
partition functions are obtained by fusion with all primaries,
starting from OO b.c.'s lends support to the conjecture that
only open and uniform fixed points occur in this problem.  Conversely,
the fact that U can be obtained from O supports the general notion
that fusion provides a complete set of fixed points starting from
a suitable reference fixed point. However, it must be admitted that
the fusion construction has not added very much to our understanding
of the SU(2) spin chain. The two basic fixed points are both
trivial and could be obtained by elementary methods.  As we shall
see in the next section, the situation is quite different in the
SU(3) case.  Now  non-trivial fixed points occurs which cannot
be obtained by elementary methods.  Fusion provides a powerful
method of solving for the properties of these fixed points.  Furthermore,
it is seems reasonable to conjecture that the set of fixed points
obtained by fusion may give the complete set of conformally invariant
boundary conditions for the model.

\section{SU(3) Case}
The Hamiltonian for the SU(3) ``spin'' chain may be written as in
Eq. (\ref{Ham1}) with the constraint of eq. (\ref{constraint}).
Alternatively, we may introduce generators of $SU(3)$, $T^A$, with
$A=1,2,3,\ldots 8$, a complete set of traceless Hermitian matrices
normalized so:
\begin{equation}
tr T^AT^B=(1/2)\delta^{AB},\end{equation}
and associated operators:
\begin{equation}
S^A_j\equiv \eta^{\dagger \alpha}_j(T^A)_\alpha^\beta\eta_{j\beta}.
\end{equation}
The Hamiltonian may then be written:
\begin{equation}
H=\sum_{jA}J_jS^A_jS^A_{j+1}.\end{equation}
The continuum limit can be derived, for example, using
a weak coupling Hubbard model representation and then extrapolating
to infinite Hubbard coupling constant.\cite{Affleck4}
  We thus keep only Fourier
modes of the fermion fields $\eta^\alpha$ near the Fermi points
$k_F=\pm \pi /3$, introducing left and right movers:
\begin{equation} \eta^\alpha_j\approx e^{i\pi j/3}\eta^\alpha_L(j)+
e^{-i\pi j/3}\eta^\alpha_R(j).\end{equation}
The resulting interacting fermion model can be treated using
non-abelian bosonization.  We thus introduce an $SU(3)_1$ WZW
non-linear $\sigma$ model field $g^\alpha_\beta$ to represent
the spin degrees of freedom and an additional charge boson field.
The charge boson develops a gap from the Hubbard interaction and
can be dropped from the Hamiltonian which is then just the
conformally invariant WZW model up to irrelevant interactions,
(and  before including impurity effects).  The original spin operators are then
represented at low energies as:
\begin{equation}
S^A_j\approx (J^A_L+J^A_R)(j) + [\hbox{constant}\cdot e^{i2\pi
j/3}tr (g(j)T^A) + h.c.],\end{equation} where $J^A_{L/R}$ are the
SU(3) currents:
\begin{equation}
J^A_{L,R}\equiv \eta^{\dagger \alpha}_{L/R}(T^A)_\alpha^\beta
\eta_{\beta L,R}.
\end{equation}
$g$ has a scaling dimension of 2/3.  Using the OPE it can be shown that:
\begin{equation}
\sum_AS^A_jS^A_{j+1}\approx [e^{i2\pi j/3}\hbox{constant}\times tr g + h.c.]
+\hbox{constant}\times \sum_AJ^A_LJ^A_R.\end{equation}
The $2k_F$ part has dimension 2/3.
s
We now consider a single modified exchange coupling $J_0=J'$
between sites 0 and 1. This introduces the interaction in the low
energy effective Hamiltonian:
\begin{equation}
\delta H \propto (J'-J) tr g + h.c.\end{equation}
Since this has dimension $2/3<1$ it is relevant.  We now consider the
possible infrared fixed point of the RG flow.  In the case $J'<J$ it
is plausible that $J'$ simply renormalizes to $0$ as in the SU(2)
case corresponding to an open chain fixed point.  However, the
situation is now quite different for $J'>J$.  Unlike the SU(2)
case, $J'\to \infty$ is {\it not} a stable fixed point.  In
the SU(3) case, if we take $J'\to \infty$ we project the two
``spins'' at sites 0 and 1 into the $\bar 3$ representation,
rather than into a singlet, as for SU(2).
Even at $J'\to \infty$
a residual interaction of $O(J)$ exists between this effective
$\bar 3$ spin and the neighbouring spins at sites (-1) and 2.

Since the sign of this residual interaction is important,
we calculate it explicitly.  This is most conveniently
done in terms of the spin operators:
\begin{equation}
S^{\alpha}_\beta \equiv \eta^{\dagger \alpha}\eta_\beta - (1/3)\delta^\alpha_\beta .
\end{equation}
This acts on the $3$ representation state,
\begin{equation}
|^\alpha> \equiv \eta^{\dagger \alpha}|0>,\end{equation}
as:
\begin{equation}
S^\alpha_\beta |^\gamma >=\delta^\gamma_\beta |^\alpha >-(1/3)\delta^\alpha_\beta |^\gamma >.
\end{equation}
The $\bar 3$ state (on a single site) corresponds to two fermions:
\begin{equation}
|^{\alpha ,\beta}>=-|^{\beta ,\alpha}>=\eta^{\dagger \alpha}\eta^{\dagger \beta}|0>.
\end{equation}
The projected $\bar 3$ state on sites 0 and 1, obtained at
 $J'\to \infty$ can be written:
\begin{equation}
|^{\alpha ,\beta}>_{01}\equiv |^\alpha >_0\times |^\beta >_1-|^\beta >_0\times |^\alpha >_1,\end{equation}
where the first and second factor refer to sites $0$ and $1$ respectively.
Now consider the action of $S^{\alpha}_{0\beta}$ on this $\bar 3$ state:
\begin{equation}
S^{\alpha}_{0\beta} |^{\gamma ,\delta}>_{01}=\delta^\gamma_\beta|^\alpha >_0
\times |^\delta >_1-\delta^\delta_\beta |^\alpha >_0\times |^\gamma >_1
-(1/3)\delta^\alpha_\beta |^{\gamma ,\delta}>_{01}.\end{equation}
Finally we project this back into the low energy $\bar 3$ subspace, by
antisymmetrizing, giving:
\begin{equation}
PS^\alpha_{0\beta}|^{\gamma ,\delta }>_{01}=
(1/2)[\delta^\gamma_\beta |^{\alpha ,\delta}>_{01}-
\delta^\delta_\beta |^{\alpha
,\gamma}>_{01}]-(1/3)\delta^\alpha_\beta |^{\gamma ,\delta}>.\end{equation}
Thus we see that, upon projecting into the low energy subspace of the $\bar 3$
representation,
\begin{equation}
{\cal P}S^{\alpha}_{0,\beta}{\cal P} = (1/2)S^\alpha_{eff,\beta}-
(1/6)\delta^\alpha_\beta .\end{equation}
Therefore, the residual exchange interaction between site (-1) and
the effective $\bar 3$ spin has the value $J/2>0$.  The case of
a small positive coupling to the effective $\bar 3$ impurity spin
gives a Kondo type RG equation and is hence marginally relevant.
Thus, it appears that $J'$ {\it does not} renormalize to $\infty$
when  $J'>J$.  It is therefore reasonable to expect that
some sort of non-trivial fixed point occurs in this problem.  We
construct this fixed point explicitly below using the fusion method.

First, however, we consider the case of two modified (but equal)
exchange couplings on neighbouring links 0-1 and 1-2.  A difference immediately
appears with the SU(2) case.  The relevant operator,
$\hbox{constant}\  tr g + h.c.$ appears in the SU(3)
case because the oscillating factors $e^{i2\pi j/3}$ do not cancel
between two neighbouring links.  Thus modifying two neighbouring
links is also a relevant perturbation.  We may consider the
possible RG flows by again considering the limit where $J'\to \infty$
or $0$.  Note that when $J'\to \infty$ , 3 neighbouring sites form
an SU(3) singlet:
\begin{equation}
\epsilon_{\alpha \beta \gamma}|^\alpha>_0\times |^\beta
>_1|^\gamma >_2. \label{singlet}\end{equation}
This effectively breaks the chain into two disconnected pieces,
corresponding to the open fixed point.  Since this is a stable
fixed point, it is plausible that it any $J'>J$ flows to it.  On
the other hand, when $J'\to 0$, we get two chains with a Kondo
coupling to an impurity in the $3$ representation.
 This corresponds to the 2-channel $SU(3)$ Kondo model, as can be
seen from the non-abelian bosonization of this model.\cite{Affleck6}
The 2 channels correspond to the decoupled chains on the two
sides of the impurity.
This Kondo interaction
is marginally relevant, so $J'=0$ is {\it not} a stable fixed point.
Thus it appears that there must be a non-trivial fixed point with
two modified links in the case $J'<J$.

We now wish to study this problem using the fusion method.  To do
this we must first introduce an appropriate conformal embedding.
Following the $SU(2)$ case, we regard the right movers as a second
branch of left movers and then introduce an $SU(3)_2$ WZW model
representing the diagonal $SU(3)$ degrees of freedom.  We then
must introduce another CFT representing the coset $SU(3)_1\times
SU(3)_1/SU(3)_2$.  This turns out to be the 3-state Potts model:
\begin{equation}
SU(3)_1\times SU(3)_2 = SU(3)_2\times \hbox{Potts}. \end{equation}
The conformal charges add up correctly: $2+2=16/5+4/5$.  The
$SU(3)_2$ WZW model has primary fields in the $3$, $6$ and $8$
representations with scaling dimensions $4/15$, $2/3$ and $3/5$
respectively (as well as the conjugate fields in the $\bar 3$ and
$\bar 6$ representations). The Potts model contains two conjugate
pairs of fields, $\sigma$, $\sigma^\dagger$ of dimension $1/15$
and $\psi$ and $\psi^\dagger$ of dimension $2/3$ as well as the
Hermitian field $\epsilon$ of dimension $2/5$. Note that we only
obtain the ``chiral factors'' of the various primary operators and
these scaling dimensions are all for these chiral factors. The
lattice ``spin'' operators are then represented in terms of these
degrees of freedom as:
\begin{equation}
S^A_j \approx J^A + \phi^A[e^{i2\pi j/3}\hbox{constant}\times \sigma + h.c.]
\end{equation}
Here $\phi^A$ is the $SU(3)_2$ adjoint representation field.  The
$2k_F$ part of the spin produce is:
\begin{equation}
\sum_AS^A_jS^A_{j+1} \approx [e^{i2\pi j/3}\times
\hbox{constant}\times  \psi +h.c.].\end{equation} We can again
write down the partition functions corresponding to trivial
uniform or open boundary conditions, more or less by inspection.
In this case we get a different result depending on the length of
the chains mod $3$, represented by superscripts $0$, $1$ or $2$.
We find that it is necessary to introduce additional characters in
the Potts sector which don't occur in the bulk Potts spectrum but
are legitimate  conformal towers occurring in the bulk spectrum of
the other c=4/5 CFT the tetra-critical Ising model.  This
phenomena was already encountered in our discussion of boundary
critical points in the Potts model.  Characters not appearing in
the bulk spectrum can occur in the spectrum with boundaries (in
the open string channel only) and can be used in constructing
boundary conditions by fusion.  These additional characters
correspond to primary fields of dimension $1/8$, $13/8$, $1/40$
and $21/40$.  The partition functions correspond to the various
uniform or open b.c.'s can then be written as follows, first in
terms of $SU(3)_1\times SU(3)_1$ characters and then in terms of
$SU(3)_2\times$ Potts characters.
\begin{eqnarray}
q^{1/6}Z^0_{UU}(q)&=&\left[ \chi_1^{(1)}(q)\right]^2
+\left[ \chi_3^{(1)}(q)\right]^2
+\left[ \chi_{\bar 3}^{(1)}(q)\right]^2=\chi_1^{(2)}(q)\left[
\chi_I^P(q)+\chi_{\psi}^P(q)+\chi_{\psi^\dagger}^P(q)\right]
\nonumber \\ &&
+\chi_8^{(2)}(q)\left[ \chi_{\epsilon}^P(q)+\chi_{\sigma}^P(q)
+\chi_{\sigma^\dagger}^P(q)\right]
\nonumber \\
q^{1/6}Z^1_{UU}(q)&=& 2\chi_1^{(1)}(q)\chi_3^{(1)}(q)+
\left[ \chi_{\bar 3}^{(1)}(q)\right]^2
=\chi_3^{(2)}(q)\left[ \chi_{\epsilon}^P(q)+
\chi_{\sigma}^P(q)+\chi_{\sigma^\dagger}^P(q)\right]\nonumber \\
&&+\chi_{\bar 6}^{(2)}(q)\left[ \chi_{I}^P(q)+
\chi_{\psi}^P(q)+\chi_{\psi^\dagger}^P(q)\right]
\nonumber \\
 q^{1/6}Z^2_{UU}(q)&=& 2\chi_1^{(1)}(q)\chi_{\bar 3}^{(1)}(q)+
\left[ \chi_{3}^{(1)}(q)\right]^2
=\chi_{\bar 3}^{(2)}(q)\left[ \chi_{\epsilon}^P(q)+
\chi_{\sigma}^P(q)+\chi_{\sigma^\dagger}^P(q)\right]\nonumber \\
&&+\chi_{6}^{(2)}(q)\left[ \chi_{I}^P(q)+
\chi_{\psi}^P(q)+\chi_{\psi^\dagger}^P(q)\right]
\nonumber \\
q^{1/6}Z^0_{UO}(q)&=&q^{1/8}\chi_1^{(1)}(\sqrt{q})
=\chi_1^{(2)}(q) \chi_{1/8}^P(q)+\chi_8^{(2)}(q) \chi_{1/40}^P(q)\nonumber \\
q^{1/6}Z^1_{UO}(q)&=&q^{1/8}\chi_3^{(1)}(\sqrt{q})
=\chi_3^{(2)}(q) \chi_{1/40}^P(q)+\chi_{\bar 6}^{(2)}(q) \chi_{1/8}^P(q)\nonumber \\
q^{1/6}Z^2_{UO}(q)&=&q^{1/8}\chi_{\bar 3}^{(1)}(\sqrt{q})
=\chi_{\bar 3}^{(2)}(q) \chi_{1/40}^P(q)+\chi_{6}^{(2)}(q) \chi_{1/8}^P(q)\nonumber \\
q^{1/6}Z^{00}_{OO}(q)&=&\left[ \chi_1^{(1)}(q)\right]^2
=\chi_1^{(2)}(q) \chi_I^P(q)+\chi_8^{(2)}(q)\chi_{\epsilon}^P(q)\nonumber \\
q^{1/6}Z^{01}_{OO}(q)&=&\chi_1^{(1)}(q)\chi_3^{(1)}(q)
=\chi_3^{(2)}(q) \chi_{\sigma}^P(q)+\chi_{\bar 6}^{(2)}(q)\chi_{\psi}^P(q)\nonumber \\
q^{1/6}Z^{02}_{OO}(q)&=&\chi_1^{(1)}(q)\chi_{\bar 3}^{(1)}(q)
=\chi_{\bar 3}^{(2)}(q) \chi_{\sigma^\dagger}^P(q)+\chi_{6}^{(2)}(q)\chi_{\psi^\dagger}^P(q)\nonumber \\
q^{1/6}Z^{11}_{OO}(q)&=&\left[\chi_3^{(1)}(q)\right]^2
=\chi_{6}^{(2)}(q) \chi_{I}^P(q)+\chi_{\bar 3}^{(2)}(q)\chi_{\epsilon}^P(q)\nonumber \\
q^{1/6}Z^{12}_{OO}(q)&=&\chi_3^{(1)}(q)\chi_{\bar 3}^{(1)}(q)
=\chi_{1}^{(2)}(q) \chi_{\psi}^P(q)+\chi_{8}^{(2)}(q)\chi_{\sigma}^P(q)\nonumber \\
q^{1/6}Z^{22}_{OO}(q)&=&\left[ \chi_{\bar 3}^{(1)}(q)\right]^2
=\chi_{\bar 6}^{(2)}(q)\chi_{I}^P(q)+\chi_{3}^{(2)}(q) \chi_{\epsilon}^P(q)
\label{trivZ}
\end{eqnarray}
Although we don't know formal proofs of these identities, we have
checked them using MATHEMATICA up to the  level $q^{40}$  in the 
expansion in $q$, in 
all cases.

 Following our treatment of the
$SU(2)$ case in the previous section, we now begin with the
open-open boundary conditions and consider the effect of all
possible fusion processes in either $SU(3)_2$ or Potts sectors.
The fusion rules for $SU(3)_2$ are:
\begin{eqnarray}
3\times 3 &\to& \bar 3 + 6\nonumber \\
3\times \bar 3 &\to & 1 + 8\nonumber \\
3\times 6 &\to & 8 \nonumber \\
3\times \bar 6 &\to & \bar 3\nonumber \\
3\times 8 &\to & 3 + \bar 6 \nonumber \\
6\times 6 &\to & \bar 6\nonumber \\
6\times 8 &\to & \bar 3\nonumber \\
6\times \bar 6 &\to & 1\nonumber \\
8\times 8 &\to & 1 + 8
\end{eqnarray}
The fusion rules, and modular S-matrix in the $W$-invariant sector of the Potts model are
equivalent to those of $SU(3)_2$ with the identification:
\begin{eqnarray}
1&\to I\nonumber \\
8 &\to \epsilon \nonumber \\
3 &\to & \sigma \nonumber \\
\bar 3 &\to & \sigma^\dagger \nonumber \\
6 &\to & \psi \nonumber \\
\bar 6 &\to & \psi^\dagger .
\end{eqnarray}
The fusion rules and modular S-matrix for the extended Potts algebra are
given in Ref. (\onlinecite{Affleck2}),
 Tables I and II.  We start with $Z^{00}_{OO}$, although we don't expect to
obtain a different result if we begin with the other possible
$Z^{ij}_{OO}$ cases.  Let us first consider fusion in the $SU(3)_2$ sector.
We see that fusion with $3$ (or equivalently $\bar 3$) or $8$ gives a
new partition function, not in the list in Eq. (\ref{trivZ}).  On the other
hand, fusion with $6$ (or equivalently $\bar 6$) gives $Z^{11}_{OO}$
(or $Z^{22}_{OO}$).  These results are more or less what we should
have expected based on the above discussion of the RG flows.  Fusion
with $3$ should correspond to weakly coupling one new impurity spin
to both chains at the origin.  This is related to weakening
two neighbouring links, in the case of antiferromagnetic coupling, which was argued above to lead to a non-trivial
 fixed point.    Fusion with $\bar 3$
would correspond to adding two impurity spins between the ends of
the open chains at the origin, with the two spins strongly coupled
together antiferromagnetically in order to obtain the $\bar 3$
representation.  This is related to the case of one strengthened
link which should also lead to a non-trivial fixed point.  Fusion
with $6$ is related to adding two impurity spins which are
ferromagnetically coupled to each other.  This is related to
weakening the (initially antiferromagnetic) coupling between two
spins which was argued to lead to a trivial fixed point, the open
chain.  The case of fusion with $8$ is less obviously related to
the previous discussion.  It is instructive, at this point to
consider a second fusion with the conjugate operator,
corresponding to the same process taking place at $x=l$.  Fusion
with $6$ then $\bar 6$ produces the open-open fixed point, as
expected.  On the other hand fusion with $3$ then $\bar 3$ or $8$
then $8$ leads to the same new partition function, which
corresponds to having the non-trivial b.c. at both ends of the
system.  The fact that the same partition function results from
either double fusion process indicates that there is only one new
fixed point occurring, not two.  Next we consider fusion in the
Potts sector.  We expect that this corresponds to adding 3
impurity spins with various types of self-couplings.  We find that
fusion with $\sigma$ (or equivalently $\sigma^\dagger$) or
$\epsilon$ leads to a non-trivial fixed point.  This appears to be
the same one obtained from $SU(3)_2$ fusion, as seen by checking
the result of double fusion.  On the other hand, fusion with
$\psi$ leads to the $Z^{12}_{OO}$ partition function, indicating
that we simply obtain a flow to open-open boundary conditions.
This seems to correspond to 3 impurity spins coupled
asymmetrically; one attaches to one chain and 2 attach to the
other.  Finally, we may consider fusion with the extended Potts
operators.  We find that fusion with the dimension $1/8$ or $13/8$
operators gives $Z^{0}_{UO}$.  This corresponds to coupling 3
impurity spins between the ends of the open chains and obtaining
the uniform fixed point.  (This is an unstable fixed point for
this process.)  Finally, fusion with the operators of dimensions
$1/40$ or $21/40$ gives another new fixed point, {\it not}
equivalent to the one discussed earlier.

The first non-trivial fixed point, discussed above, can be given
an interpretation related to the Potts model.  In the Potts model
there is a ``mixed'' fixed point corresponding to the Potts variables
on the boundary fluctuating back and forth between two of the
three possible states.  We may interpret these 3 Potts states
as corresponding to the three possible trimerization patterns
of the $SU(3)$ spin chain.  i.e. the trimers form on sites
$3j-(3j+1)-(3j+2)$ or $(3j-1)-3j-(3j+1)$ or $(3j-2)-(3j-1)-3j$
for all integer $j$.  Note that two strengthened neighbouring
links on links 0-1 and 1-2 favor trimer formation on 0-1-2,
corresponding to a fixed b.c. in the Potts model and an open
b.c. in the spin chain.  Similarly one weakened bond on 0-1
favors the  $1-2-3$ trimerization pattern, again
corresponding to the open b.c.  However, one strengthened
bond on 0-1 equally favors two trimerization patterns,
0-1-2 or (-1)-0-1.  We may think of the non-trivial fixed point
as being one in which the trimers resonant between these
two states near the origin.  This is very analogous to the
mixed fixed point in the Potts model.  Hence it is appropriate
to refer to this state in the $SU(3)$ chain as the mixed
fixed point.  Similarly, weakening two bonds on 0-1 and 1-2
equally favors two trimerization patterns 1-2-3 or (-1)-0-1.
Again this gives the mixed fixed point.  We note that fusion
in the Potts sector of the $SU(3)$ chain connects the fixed
points in a way which corresponds to that in the Potts model.
Starting from a fixed b.c. in the Potts model, fusion
with $\psi$ (or $\psi^\dagger$) gives the other fixed b.c.'s
but fusion with $\sigma$, $\sigma^\dagger$ or $\epsilon$
gives the mixed b.c.  Fusion with the $1/8$ operator
gives the free b.c. in the Potts model corresponding to
the uniform b.c. in the $SU(3)$ chain.  Fusion with
the $1/40$ operator in the Potts model gives a new fixed
point first discussed in Ref. (\onlinecite{Affleck2}).  This
corresponds to the new fixed point in the $SU(3)$ chain.  So
far, we have been unable to understand what sort of microscopic
impurity couplings in the $SU(3)$ chain would produce a flow
to this new fixed point.  A related difficulty, is that the
 $Z_3$ symmetry in the impurity problems  may only be defined
in the low energy continuum limit.  While this symmetry can be
identified as translations by 0, 1 or 2 sites in the uniform
chain, this translational symmetry is always broken in the
impurity models.  Note that we were able to circumvent the
analogous problem in the case of the $SU(2)$ chain with an
impurity by identifying the $Z_2$ symmetry with reflection about a
site, rather than translation by one site.  Both these symmetry
operations have the effect of interchanging the two dimerized
groundstates.  On the other hand, in the $SU(3)$ case there appear
to be no analogous symmetries which interchange the trimerized
groundstates and which remain symmetries with impurity
interactions present.

It is instructive to consider the boundary operator content at
the mixed fixed point, obtained by double fusion.  This is:
\begin{equation}
1\times I + 8\times I + (2\times )8\times \epsilon + 1\times
\epsilon.\label{opcontent} \end{equation} Here the first and
second factors correspond to operators in the $SU(3)_2$ and Potts
sector respectively and the factor of 2 indicates that two such
operators occur.  We see that there is only 1 $SU(3)$ symmetric
relevant  operator, $1\times \epsilon$.  This has a natural
interpretation related to the above discussion.  The mixed fixed
point corresponds to resonance between two different trimerization
patterns near the origin.  Modifying exchange couplings so as to
favor one of these over the other is a relevant perturbation.  For
instance, if we obtain the mixed fixed point by strengthening the
coupling on link 0-1, then strengthening the coupling on 1-2 is
relevant since it then favors the 0-1-2 trimerization over
(-1)-0-1.    In fact, the corresponding relevant operator also
occurs at the mixed fixed point in the Potts model.

\section{Connection with $\lowercase{t}JV$ model}
The SU(3) ``spin'' chain is equivalent to the $tJV$ model for a
special choice of the parameters $J$ and $V$.\cite{Lai}  The Hamiltonian
is:
\begin{equation}
H=\sum_j\left\{ [-tP\psi^{\dagger a}_j\psi_{j+1,a}P + h.c.]
+J\vec S_j\cdot \vec S_{j+1}+Vn_jn_{j+1}\right\} . \label{tJV}
\end{equation}
Here $P$ projects out states with no double occupancy, and $\vec S_j$
and $n_j$ are the electron spin and charge operators on site $j$:
\begin{eqnarray}
\vec S_j &\equiv & \psi^{\dagger a}_j{\vec \sigma^b_a \over 2}\psi_{jb}
\nonumber \\
n_j &\equiv & \psi^{\dagger a}_j\psi_{ja}.
\end{eqnarray}
The spin indices, $a$, $b$, represented by Latin letters,
are summed from $1$ to $2$ only.  Greek letters
are used for the $SU(3)$ indices summed over $0$, $1$, $2$.
This model is equivalent to the $SU(3)$ spin chain for $J=2t$, $V=3t/2$.

We remark that this {\it is not} the same as the supersymmetric $tJ$
model which has $J=2t$ but $V=-t/2=-J/4$.  (The latter is referred to as
a $tJ$ model rather than a $tJV$ model because the $tJ$ model
is sometimes written as in Eq. (\ref{tJV}) with
$V=-J/4$.)

The equivalence of the $SU(3)$ spin chain with
this $tJV$ model is established by writing the $SU(3)$ spin
chain Hamiltonian in the form:
\begin{equation}
H=(J/2) \sum_jS^{\alpha}_{j\beta}S^{\beta}_{j\alpha}\label{SU3Ham}
\end{equation}
with
\begin{equation}
S^\alpha_{j\beta} \equiv \eta^{\dagger
\alpha}_j\eta_{j\beta},\end{equation}
 and then mapping into
 $tJV$ operators as follows:
 \begin{eqnarray}
 \eta^{\dagger a}_j\eta_{jb}&\to& \psi^{\dagger a}_j\psi_{jb}\nonumber
 \\
 \eta^{\dagger 0}_j\eta_{j0} =1-\eta^{\dagger a}_j\eta_{ja}
  &\to & 1-\psi^{\dagger a}_j\psi_{ja}\equiv 1-n_j\nonumber \\
  \eta^{\dagger a}_j\eta_{j0} &\to & \exp \left[ i\pi
  \sum_{l=1}^{j-1}n_l\right] \psi^{\dagger a}_j
  (1-n_{j\check a})\label{mapping}
  \end{eqnarray}
  Here  $\check a$ denotes the other index.  i.e. $\check 1=2$ and
  $\check 2=1$.
  Note the familiar Jordan-Wigner string operator in the last line of
  Eq. (\ref{mapping})
   which turns the right hand side into a commuting (bosonic) object. It
   may be verified that the mapping of Eq. (\ref{mapping}) respects the
   $SU(3)$ commutation relations:
   \begin{equation}
   [S^{\alpha}_{j\beta},S^{\gamma}_{k\epsilon}]=\delta_{jk}
   [\delta_\beta^\gamma
   S^{\alpha}_{j\epsilon}-\delta_\epsilon^\alpha S^{\gamma}_{j\beta}].
   \end{equation}
   Using Eq. (\ref{mapping}), it is straightforward to check that
    the Hamiltonian of Eq. (\ref{SU3Ham})  maps into the
   $tJV$
   Hamiltonian with $t=1$, $J=2$ and $V=3/2$.  If the $SU(3)$
   Hamiltonian
   is written with periodic b.c.'s then we obtain the $tJV$ model with
   b.c.'s that are periodic  when the total number of electrons is even
   but anti-periodic when the total number is odd.  Furthermore, the
   density of electrons must be fixed at 2/3 (and the total spin at 0)
   to correspond to the $SU(3)$ invariant groundstate.

   It is also interesting to consider the continuum limit of the $SU(3)$
   spin chain using Abelian bosonization, rather than the non-abelian
   bosonization used in the previous section. This amounts to
   representing the $SU(3)_1$ WZW model in terms of two free bosons,
   a correspondence which is consistent with the central charge $c=2$.
   The two bosons can then be identified with the charge and spin
   bosons that are familiar in the continuum limit of the
   Hubbard or $tJV$ models.  We can then determine the compactification
   radii of the charge boson as well as that of the spin boson at
   the $SU(3)$ invariant point.  [The result for the spin boson
   is the well-known value corresponding to $SU(2)$ invariance.]
   At these special radii, the continuum 2-boson model becomes
   equivalent to the $SU(3)_1$ WZW model.

   The abelian bosonization of the continuum limit field theory for
the $\eta$ fermions introduces three boson, $\phi_\alpha$ for the
3 fermion fields.  These may be rewritten in terms of a more
convenient basis: $\phi$, $\phi_c$ and $\phi_s$, defined as:
\begin{eqnarray}
\phi &\equiv & {\phi_1+\phi_2+\phi_0\over \sqrt{3}}\nonumber \\
\phi_c &\equiv & {\phi_1+\phi_2-2\phi_0\over \sqrt{6}}\nonumber \\
\phi_s &\equiv & {\phi_1-\phi_2\over \sqrt{2}}.
\end{eqnarray}
$\phi$ represents the pseudo-charge boson in the 3-component $\eta$
field theory.  The Hubbard interaction in the $\eta$ theory
produces a gap for $\phi$ which may be dropped from the Hamiltonian.
$\phi_c$ is the charge boson in the 2-component, $\psi$ field
theory and $\phi_s$ is the $SU(2)$ spin boson. To determine the
radius of the charge boson we may write down the resulting
bosonized form for various operators and compare to the standard
result.  For instance:
\begin{equation}
\psi^{\dagger}_{L1}\psi_{R1}\propto e^{i\sqrt{4\pi}\phi_1}
\to e^{i[\sqrt{4\pi /3}\phi + \sqrt{2\pi /3}\phi_c
+\sqrt{2\pi}\phi_s]}.\label{abbos}
\end{equation}
We expect the Hubbard interaction to produce a non-zero
expectation value:
\begin{equation}
<e^{i\sqrt{4\pi /3}\phi}>\neq 0,
\end{equation}
so we may drop the first term in the exponent in Eq. (\ref{abbos}).
On the other hand, if we had started from the $\psi$ fermion
model (the ordinary 2-component Hubbard model) we would have
written:
\begin{equation}
\psi^{\dagger}_{L1}\psi_{R1}
\propto e^{i[\phi_c/R_c+\phi_s/R_s]}.
\label{RcRsdef} \end{equation}
In the non-interacting limit $R_c=R_s=1/\sqrt{2\pi}$.  $SU(2)$
invariant interactions renormalize $R_c$, but not $R_s$.
Comparing Eq. (\ref{RcRsdef}) to (\ref{abbos})  we see that,
the $SU(3)_1$ WZW model, the continuum limit of the $SU(3)$ invariant
$tJV$ model, has:
\begin{equation}
R_c=\sqrt{3/2\pi}.\end{equation}
In the notation of Kane and Fisher, this corresponds to:
\begin{equation}
g_{\sigma}=2,\ \  g_{\rho} = 2/3.\end{equation}
At this value of $R_c$, $\psi^{\dagger}_{L1}\psi_{R1}$ has
scaling dimension 2/3, corresponding to the 11 component of
the $SU(3)_1$ WZW field, $g_{11}$.  Here we have used
the fact that the $SU(3)$ symmetry protects the  radius
of the charge (as well as spin) boson from renormalizing
as the Hubbard interaction is increased to $\infty$.  Thus
we see that the spinful Luttinger liquid model
 at the special values of $g_{\rho}$ and $g_{\sigma}$
where it was solved exactly by Yi and Kane, has a hidden
$SU(3)$ symmetry.  This provides some understanding of
the solvability at this special point and suggests
that the $SU(3)$ approach used here is the most natural
way of studying the problem.

\section{SU(3) symmetry breaking}
A natural question to ask, at this point, is what happens
if we allow boundary interactions that break the $SU(3)$
symmetry down to $SU(2)\times U(1)$?  This would correspond
to starting with the $SU(3)$ invariant $tJV$ model in
the bulk but then allowing arbitrary strength hopping,
spin exchange and Coulomb repulsion on the modified links,
for example. Alternatively, we could apply an $SU(3)$ ``field''
which favors holes over electrons at one or more sites.
 Referring to Eq. (\ref{opcontent}), we
see that in addition to the $SU(3)$ invariant relevant operator at
the mixed fixed point there is also one relevant operator
transforming under the adjoint representation of $SU(3)$.  This
contains one $SU(2)\times U(1)$ singlet field, the 3-3 component
of the adjoint field, $\phi_{33}$. There are also two marginal
operators with the same $SU(3)$ transformation properties.  We
expect that these correspond to interactions that break both
parity and $SU(3)$.  This follows from assigning an odd parity
quantum number to the Potts operator $\epsilon$ which appears as a
factor in these marginal operators.  It is interesting to consider
what happens in the case of two weakened links, on (-1)-0 and 0-1,
 if we then apply a ``field'' (i.e.
local potential) at the origin, thus respecting
the parity symmetry.  It seems plausible that this
is a relevant perturbation, and generates the operator
$\phi_{33}$. It is clear that a large local potential at
the origin will lead to a trivial fixed point.  A
positive potential, favoring a hole at the origin gives the
open fixed point, corresponding to zero conductance for
spin and charge.  On the other hand, a negative potential,
favoring one electron at the origin, produces a trivial,
but non $SU(3)$-invariant fixed point, as argued by
Kane and Fisher.\cite{Kane}  The single electron at the origin
acts as a Kondo impurity.  Since $g_{\rho}<1$, it blocks charge transport
through the origin but allows spin transport.  This corresponds
to a phase with perfect transmission for spin but perfect
reflection for charge. The non-trivial, mixed,
critical point is sitting at a resonance, in between these
two stable fixed points.  Kane and Fisher identified
their non-trivial resonant fixed point as an unstable fixed point
 separating precisely these two stable phases.
 Remarkably, $SU(3)$ symmetry (at the boundary as well as in the bulk)
 puts the system exactly on resonance.

We obtain an analogous result by considering the case of
one strengthened link with breaking of $SU(3)$ symmetry.  It
is instructive to consider the effect of the $SU(3)$ symmetry
breaking in the limit where $J'\to \infty$.  Then, the
$\bar 3$ representation is projected out on the sites 0-1.
The 3 states of the $\bar 3$ representation consist of an
$SU(2)$ doublet, with one electron (of either spin) hopping
back and forth between sites 0 and 1 in a zero momentum state,
 and of an $SU(2)$ singlet state with 1 electron on site 0
and 1 electron on site 1. In other words, we may either
increase $t$ to favor the doublet state or increase $J$ to
favor the singlet.
 $SU(3)$ symmetry breaking favors
either the doublet or the singlet.  In the case where the
singlet is favored we again get the open fixed point since
there is zero charge or spin transport through sites 0-1
in this case.  However, when the doublet is favored the
single electron shared by sites 0 and 1 again acts
 like a Kondo impurity, allowing spin transport but
not charge transport.  Again we expect flow to a fixed
point with perfect transmission for spin but perfect
reflection for charge.

We also consider breaking of the $SU(3)$ symmetry down to
$SU(2)\times U(1)$ in the bulk.  This corresponds to the $tJV$
model with general parameters and chemical potential.  The
relevant boundary operator, $\phi_{33}$ discussed in the previous
paragraph will remain relevant for a range of bulk anisotropy
(although with anisotropy-dependent scaling dimension) and will
generally be present in the effective Hamiltonian, unless
fine-tuning is done. The stable fixed points are the two trivial
ones  discussed in the previous paragraph.  A non-trivial fixed
point appears as an unstable ``resonance'' critical point. The
critical exponents at this non-trivial critical point should vary
continuously with bulk anisotropy. However, they are only known at
the $SU(3)$ symmetric point (and at two other points where the
non-trivial critical point merges with one of the trivial critical
points, using the ``$\epsilon$'' expansion\cite{Kane}).  Thus we
see that the $SU(3)$ invariant spin chain has the very special
property that the non-trivial critical point is stabilized by a
symmetry. This $SU(3)$ invariant spin chain (or equivalently $tJV$
model) would thus provide a convenient model for numerical study
of this non-trivial critical point.

It is also interesting to consider a different type of $SU(3)$
symmetry breaking: $SU(3)\to SO(3)$ such that the $3$ rep of
$SU(3)$ transforms under the triplet ($j=1)$ rep of SO(3).  As
shown by Itoi and Kato,\cite{Itoi} this symmetry breaking pattern occurs in
ordinary SU(2) spin-1 spin chains with biquadratic as well as
bilinear exchange interactions:
\begin{equation}
H=J\sum_j[\cos \theta \vec S_j\cdot \vec S_{j+1}+ \sin \theta
(\vec S_j\cdot \vec
S_{j+1})^2].\end{equation} The model with $\theta = \pi /4$
is exactly equivalent to the $SU(3)$ spin chain.  Varying $\theta$,
corresponds to this pattern of $SU(3)$ symmetry breaking in the
continuum limit $SU(3)_1$ WZW model.  Only marginal symmetry
breaking  interactions are generated in the effective Hamiltonian.
In the case $\theta > \pi /4$ these can be shown to be marginally
irrelevant.\cite{Itoi} The remarkable conclusion is that the S=1
chain has a gapless phase for all $\pi /4 <\theta < \pi /2$.  (On the other
hand, for $-\pi /4 < \theta < \pi /4$, the system goes into the Haldane
gap phase.)
 The effective Hamiltonian of the gapless phase is the $SU(3)_1$
 WZW model, up to logarithmic symmetry-breaking
 corrections.
Now let us consider the effect of this pattern of SU(3) symmetry
breaking in the impurity models.  Noting that the 8 rep of $SU(3)$
decomposes into the direct sum of spin $j=2$ and $j=1$ reps, with
no $SO(3)$ singlets, we conclude that no relevant or marginal
operators are allowed in the effective boundary Hamiltonian at the
mixed critical point, even when the $SU(3)$ symmetry is broken
down to $SO(3)$.  Thus the boundary critical phenomena that we
have elucidated for the $SU(3)$ invariant model should also occur
in the general bilinear-biquadratic spin-1 chain, with bulk
couplings $\pi /4<\theta < \pi /2$.  This holds out the possibility of
experimental observation of these critical phenomena.

\section{Connection with Kondo model and quantum Brownian motion and extension to SU(n)}
There is clearly a close connection between the RG flows
 that we have discussed in the $SU(3)$ spin chain and those in
 the $SU(3)$ 2-channel Kondo model.  As already mentioned
 below Eq. (\ref{singlet}), starting from the case of
 two equal weak links is equivalent to the RG flow from
 weak coupling in the Kondo model, with the 2 decoupled
 chains on either side of the central spin acting as the
 2 channels.  In the continuum limit the correspondence is
 also clear from the occurrence of the $SU(3)_2$ WZW model.
 The mixed critical point in the spin chain corresponds to
 the non-trivial overscreened fixed point in the Kondo model.
 [For a discussion of this model see Refs. (\onlinecite{Parcollet,Jerez}).] 
 In both
 models this fixed point can be obtained by fusion with
 the $3$ representation operator in the $SU(3)_2$ WZW model.
 However, the phase diagram at stronger coupling (beyond
  the non-trivial critical point) is different in the two
  models.  In the spin chain, at stronger coupling we
  encounter the (unstable) uniform fixed point and then
  at infinite coupling the open fixed point.  On the
  other hand in the Kondo model, the only other fixed point
  is expected to be the unstable overscreened one at
  infinite coupling.  We also remark that breaking the reflection symmetry, 
so that one weak link is of different strength than the other, is 
equivalent to channel symmetry breaking in the 2-channel Kondo model.  This
is a relevant perturbation at the non-trivial fixed point in both cases. 

  It is also worth remarking in more detail on the connection
  between the boundary critical phenomena that we have
  been discussing in the $SU(3)$ spin chain and that in
  the model of quantum Brownian motion (QBM) in Ref.
  (\onlinecite{Affleck1}).  The latter  model has two massless
  bosons, defined on the half-line, with boundary sine-Gordon
  interactions.  The $SU(3)$ spin chain
   can be regarded as having 4 left-moving massless bosons
  on the half-line, corresponding to the central charge 4
  in Eq. (\ref{2+2}).  The $SU(3)_2$ WZW model can be written
  as a conformal embedding of 2 free bosons [corresponding
  to the maximal abelian subgroup of $SU(3)$] together with
  a $c=6/5$ conformal field theory which can apparently be
  regarded as the $Z_3^{(5)}$ CFT discussed in Ref.
  (\onlinecite{Affleck1}).  This, together with the Potts model
  ($c=4/5$) comprise the 2 free bosons ($c=2$) occuring
  in the QBM model.  Since the extra 2 free bosons of the
  spin chain don't occur in the effective Hamiltonian if
  $SU(3)$ symmetry [or even its $U(1)\times U(1)$ subgroup]
  is maintained, there is a correspondence between RG
  flows in QBM and in the spin chain.  The Dirichlet, Neumann,
  $Y$ and $W$ fixed points in the QBM model correspond
  respectively to open, uniform, mixed and new fixed points in
  the spin chain.
  In both models all these fixed points can be constructed
  by fusion with the same operators in the Potts sector
  starting from the Dirichlet (i.e. open) fixed point.

Finally, we remark that most of the considerations of this
paper can be extended to the general case of $SU(n)$ ``spin'' chains.
After regarding the right-movers as a second branch of left-movers, 
we can again introduce a conformal embedding:
\begin{equation}
SU(n)_1\times U(n)_1 = SU(n)_2\times Z_n,\end{equation}
where $Z_n$ refers to the $Z_n$ parafermion conformal field theory. 
Non-trivial critical points can again be constructed by the fusion 
method and given a physical interpretation in these lattice models.
These fixed points have already been discussed in the context of 
quantum Brownian motion.\cite{Yi,Affleck1}

\acknowledgements
Much of this research was carried out while all three authors were 
visiting members of the Institute for Theoretical Physics in Santa 
Barbara, in July, 1997.  This research was supported in part by
the NSF grant No. PHY-94-07194 at the ITP, by NSERC of Canada (IA), 
by a  Grant-in-Aid
from Ministry of Education, Science, Sports and Culture
of Japan (M.O.), and by the DOE and
the NSF under the NYI program (H.S.).

\end{document}